\documentclass[useAMS,usenatbib]{mn2e}
\usepackage{mn2e-breakabs}
\usepackage{graphicx}
\usepackage{times}
\usepackage{float}
\usepackage{hyperref}
\usepackage{subfig}
\usepackage{amsmath}
\def\la{\mathrel{\mathpalette\fun <}}
\def\ga{\mathrel{\mathpalette\fun >}}
\def\fun#1#2{\lower3.6pt\vbox{\baselineskip0pt\lineskip.9pt
        \ialign{$\mathsurround=0pt#1\hfill##\hfil$\crcr#2\crcr\sim\crcr}}}

\newcommand{\be}{\begin{equation}}
\newcommand{\ee}{\end{equation}}
\newcommand{\ba}{\begin{eqnarray}}
\newcommand{\ea}{\end{eqnarray}}
\newcommand{\simgt}{\,\hbox{\lower0.6ex\hbox{$\sim$}\llap{\raise0.6ex\hbox{$>$}}}\,}
\newcommand{\simlt}{\,\hbox{\lower0.6ex\hbox{$\sim$}\llap{\raise0.6ex\hbox{$<$}}}\,}

\begin{document}

\title
{A closer look at the cosmological implications of the $\Lambda$HDE model}

\author[Shuang Wang et al.]
{
  \parbox{\textwidth}{
    Shuang Wang$^{1}$\thanks{wangshuang@mail.sysu.edu.cn}
    Sixiang Wen$^{1}$,
    Lanjun Zhou$^{2,3}$,
    Miao Li$^{1}$
}
  \vspace*{4pt} \\
$^1$ School of Physics and Astronomy, Sun Yat-Sen University, Zhuhai 519082, P. R. China\\
$^2$ Key Laboratory of Theoretical Physics, Institute of Theoretical Physics, Chinese Academy of Sciences, Beijing 100190, P. R. China\\
$^3$ School of Physical Science, University of Chinese Academy of Science, Beijing 100049, P. R. China\\
}

\date{\today}

\maketitle

\begin{abstract}

In a previous paper \cite{Hu2015}, we proposed a heterotic dark energy model, called $\Lambda$HDE, in which dark energy is composed of two components: cosmological constant (CC) and holographic dark energy (HDE). The aim of this work is to give a more comprehensive and systematic investigation on the cosmological implications of the $\Lambda$HDE model. Firstly, we make use of the current observations to constrain the $\Lambda$HDE model, and compare its cosmology-fit results with the results of the $\Lambda$CDM and the HDE model. Then, by combining a qualitative theoretical analysis with a quantitative numerical study, we discuss the impact of considering curvature on the cosmic evolutions of fractional HDE density $\Omega_{hde}$ and fractional CC density $\Omega_{\Lambda}$, as well as on the ultimate cosmic fate. Finally, we explore the effects of adopting different types of observational data. We find that: (1) the current observational data cannot distinguish the $\Lambda$HDE model from the $\Lambda$CDM and the HDE model;
this indicates that DE may contain multiple components.
(2) the asymptotic solution of $\Omega_{hde}$ and the corresponding cosmic fate in a flat universe can be extended to the case of a non-flat universe; moreover, compared with the case of a flat universe, considering curvature will make HDE closer to a phantom dark energy.
(3) compared with JLA dataset, SNLS3 data more favor a phantom type HDE;
in contrast, using other types of observational data have no significant impact on the cosmic evolutions of the $\Lambda$HDE model.

\end{abstract}


\section{Introduction}
\label{sec:intro}

Since its discovery in 1998~\cite{Riess1998,Perl1999}, dark energy (DE) has
become one of the central problems in modern cosmology~\cite{Frieman08,Caldwell2009,Li11,Li13,WWL16}.
Although numerous DE models \cite{Steinhardt1999,Zlatev99,Armendariz1999,
Kamenshchik01,Bento2002,Caldwell2002,MCLT03,Wei05,Cai07,ZXW07,Wang08a,Wang08b,Wei09,Gao09,Huang09,WLL11}
have been proposed, the nature of DE is still a mystery.

In principle, the DE problem should be an issue of quantum gravity \cite{Witten00}.
It is commonly believed that the holographic principle \cite{Hooft93, Susskind95} is
a fundamental principle of quantum gravity.
Based on the holographic principle,
one of the present authors (ML) proposed the holographic dark energy (HDE) model \cite{Li2004}.
In this model, the HDE density $\rho_{hde}$ can be described by
\begin{equation}
  \label{defHDE}
  \rho_{hde}=3C^2M_p^2L^{-2},
\end{equation}
where $C$ is a key parameter to label the magnitude of holographic contribution to DE,
$M^2_p = (8\pi G)^{-1}$ is the reduced Planck mass, and $G$ is Newtonian gravitational constant.
Li suggested to choose the future event horizon ($R_h=a\int_t^{+\infty}\frac{dt}{a}$) as the cutoff $L$ \cite{Li2004}
\footnote{The original HDE model have the circular logic problem.
In other words, the existence of the future event horizon
needs the accelerating expansion, while the dark energy component that cause accelerating expansion
is given by the future event horizon. Due to this problem, the original HDE model cannot be derived by
the action principle. But in a paper of our group \cite{Li2012}, we have constructed a action,
which can derive the energy density form of HDE from the first principle. This means that, although it
seems that the original HDE model has the circular logic problem, this problem is not a real problem
in the framework of HDE.} .
This model is the first DE model inspired by the holographic principle,
and it is in good agreement with the current cosmological observations
\cite{Huang2004, SBER05, Chang06, Zhang2007, LLWZ09, LLWWZ09, ZLLWZ12, Li2013, LWLZ13, LLMZZ13}.
In recent years, the HDE model has attracted lots of interests
\cite{Horvat04,WGA05, Pavon05, Nojiri06, Setare06, Setare07, Saridakis07}.
Our group has also done a series of research works about HDE.
For examples, in \cite{Li2008}, the HDE model was proved to be a perturbatively stable model;
in \cite{LLWZ09}, it was found that the original HDE model is more favored by the current
observations than other types of holographic DE models \cite{Cai07,Gao09};
in \cite{Li2010}, it was proved that the Casimir energy
of the photon field in a de Sitter space have the same form of HDE;
in \cite{LW10}, the entropic force formalism was applied to cosmology,
which cause that the HDE appear in the Friedmann equation.

In almost all the DE models, DE contains only a single component. However,
since baryonic matter contains multiple components, while dark matter may also contain
multiple components \cite{Bertone}, it will be very interesting to take into account
the possibility that DE contains multiple components.
In the recent paper \cite{Hu2015}, we
proposed the $\Lambda$HDE model \footnote{There is an implicit
assumption in the $\Lambda$HDE model that once the accelerated expansion commences it will never
end. It cannot be excluded that the quantum vacuum has
the possibility of decaying into radiation and/or matter particles
\cite{Freese87,Polyakov10}. However, in this work, we assume that the vacuum will not decay.},
which contains two DE components:
cosmological constant (CC) and HDE. So far as we know, this is the first theoretical attempt
to explore the possibility that DE may contain multiple components.
In \cite{Hu2015},
we simply discussed the cosmic evolutions of the $\Lambda$HDE model in a flat universe,
and constrain this model with some observational data.
However, it must be pointed out that there are still some shortcomings for Ref. \cite{Hu2015}:
(1) the cosmology-fit results of the $\Lambda$HDE model have not been compared with the
results of other DE models;
(2) the impact of curvature on the cosmic evolutions and the cosmic fates of the $\Lambda$HDE model
have not been considered;
(3) the effects of adopting different types of observational data have not been taken into account.

In this work, we will make a more comprehensive and systematic investigation on the $\Lambda$HDE model.
Firstly, we make use of a combination of the type Ia supernovae (SNe Ia), the Baryon Acoustic Oscillations (BAO),
the Cosmic Microwave Background (CMB) and the Growth Fact (GF) data to
constrain the $\Lambda$HDE model, and then compare its cosmology-fit results
with the results of the $\Lambda$-cold-dark-matter ($\Lambda$CDM) model and the HDE model.
Secondly, we extend the discussions about the cosmic evolution of $\Omega_{hde}$ in a
flat universe to the case of a non-flat universe.
We not only perform a qualitatively analysis on the evolution of $\Omega_{hde}$,
but also give a quantitative result by using the current cosmological observations.
Thirdly, in order to explore the impact of
different datasets on the evolutions of DE, we adopt two SN datasets, two
BAO datasets and two CMB datasets; in addition, we also study the
impact of adding growth factor data or not.

Different from Ref. \cite{Hu2015}, in this work we study a new version of $\Lambda$HDE model:
in the deceleration expansion stage, DE only contains the CC term; in the accelerated expansion stage, DE contains both the CC and the HDE.
We describe our method in section \ref{sec:methodology}, present our results in section
\ref{sec:results}, and summarize in section \ref{sec:conclusion}.
In this paper, we assume
today's scale factor $a_0=1$, thus the redshift $z=a^{-1}-1$. The subscript ``0'' always indicates
the present value of the corresponding quantity.

\section{Methodology}
\label{sec:methodology}
In this section, firstly we review the theoretical framework of the $\Lambda$HDE model,
then we briefly describe the observational data used in the present work.

\subsection{The $\Lambda$HDE model}
In a non-flat universe, the Friedmann equation can be written as
\begin{equation} \label{f.e.}
    3M_{p}^{2}H^{2}=\rho_{m}+\rho_{de}+\rho_{r}+\rho_{k},
\end{equation}
where $H \equiv \dot{a}/a$ is the Hubble parameter (the dot denotes the
derivative with respect to time $t$), $\rho_{m}$, $\rho_{de}$, $\rho_{k}$
and $\rho_{r}$ are the energy densities of matter, DE, curvature and
radiation, respectively. Note that $\rho_m$ is the sum of
baryon density $\rho_{b}$ and dark matter density $\rho_{dm}$.
For convenience, we define the fractional energy density of each component as:
\begin{equation}
  \Omega_{x}\equiv{\rho_{x} \over \rho_{c}},\ \ \ \
  \rho_{c}={\rho_{m}+\rho_{de}+\rho_{r}+\rho_{k}},
\end{equation}
where the subscript $x$ can represent arbitrary cosmological component.

As mentioned above, in this work we study a modified version of $\Lambda$HDE model:
in the deceleration expansion stage, DE only contains the CC term; in the accelerated expansion stage, DE contains both the CC and the HDE.
The latest Planck 2015 paper \cite{Planck201513} gives a best-fit value $\Omega_{m0}=0.308$;
this implies that the expansion of our universe begin to accelerate at $z\simeq0.65$.
Thus the total energy density can be written as
\begin{eqnarray}
  \rho_{de} &=& \left\{ \begin{array}{ll}
  \rho_{\Lambda}, & \textrm{$z>0.65$}\\
  \rho_{\Lambda}+\rho_{hde},& \textrm{$z\leq0.65$}
\end{array} \right.
\end{eqnarray}
Here $\rho_{\Lambda} = M_p^2 \Lambda$ and $\rho_{hde} = 3C^2M_p^2/L^2$ are the energy density of CC and HDE, respectively.
The cutoff length scale $L$ takes the form \cite{HuangLi2004}
\begin{equation}
  L=\frac{a}{\sqrt{|\Omega_{k0}|}H_0} sinn(\sqrt{|\Omega_{k0}|}H_0 \int_t^{+\infty} \frac{dt'}{a}),
  \label{eq:L}
\end{equation}
where the function $sinn(x)$ is defined as
\begin{eqnarray}
  sinn(x) &=& \left\{ \begin{array}{ll}
  sin(x), & \textrm{if $\Omega_{k0}<0$}\\
  x,      & \textrm{if $\Omega_{k0}=0$}\\
  sinh(x),& \textrm{if $\Omega_{k0}>0$}
\end{array} \right.
\end{eqnarray}

Let us focus on the cosmic evolution of the $\Lambda$HDE model at the redshift region $z\leq0.65$.
Following the procedure of \cite{Hu2015}, we can get the following differential
equations for the reduced Hubble parameter $E(z)\equiv H(z)/H_{0}$ and the fractional HDE
density $\Omega_{hde}(z)$:
\begin{eqnarray}
  \label{eq:OH3}
  {1\over E(z)}{dE(z) \over dz}&=&-{\Omega_{hde}\over
  1+z}\left({3\Omega_{\Lambda}+\Omega_k-\Omega_r-3\over2\Omega_{hde}}+{1\over2}+\right. \nonumber \\
    & &+\left. \sqrt{{\Omega_{hde}\over C^2}+\Omega_k} \right), \\
    {d\Omega_{hde}(z)\over dz}&=&
  -{2\Omega_{hde}(1-\Omega_{hde})\over 1+z}\left(\sqrt{{\Omega_{hde}\over
  C^2}+\Omega_k}+{1\over2}-\right. \nonumber \\
 & &\left. -{3\Omega_{\Lambda}+\Omega_k-\Omega_r\over 2(1-\Omega_{hde})}\right).
  \label{eq:omhde}
\end{eqnarray}
Making use of the initial condition $E(0)=1$ and
$\Omega_{hde}(z=0)=\Omega_{hde0}$, the Eqs. \ref{eq:OH3} and \ref{eq:omhde} can
be solved numerically.

\subsection{Observational Data}

We will make use of the following datasets to constrain the $\Lambda$HDE model.
\begin{itemize}
\item
The SNLS3 combined sample (hereafter SNLS3) consists of 472 SNe Ia \cite{Conley2011}.
This sample has been widely used in the studies of cosmology \cite{LLWZHL11, LWHZL12, HLLW16CA}.
Its $\chi^2$ function can be written as:
\begin{equation}
\chi_{SNLS3}^2 = \Delta\mbox{\bf m}^T \cdot \mbox{\bf Cov}^{-1} \cdot \Delta\mbox{\bf m},
\end{equation}
where $\mbox{\bf Cov}$ is the total covariance matrix, which is given by \cite{Conley2011},
and $\Delta \mbox{\bf m} \equiv \mbox{\bf m}_{B}-\mbox{\bf m}_{mod}$ is a data
vector. Here $\mbox{\bf m}_{B}$ is the observed peak magnitude in rest-frame $B$
band, $\mbox{\bf m}_{mod}$ is predicted magnitude of SNe Ia, given by
\begin{equation}
  \mbox{\bf m}_{mod}=5 \log_{10}[\frac{d_L}{Mpc}] + 25 - \alpha \times (s-1) +\beta \times {\cal C} + M,
\end{equation}
where $s$ describes the time stretching of the light-curve,
${\cal C}$ describes the supernova color at maximum brightness,
$M$ is the absolute magnitude, and $d_L$ is the luminosity distance
(the corresponding expression is given in \cite{WangWang13}).
Notice that the stretch-luminosity parameter $\alpha$ and the color-luminosity
parameter $\beta$ are free parameters
\footnote{It should be mentioned that, previous studies on the SNLS3 dataset
\cite{WangWang13} and JLA dataset \cite{LLWZ16} found strong evidence for the
redshift-dependence of color-luminosity parameter $\beta$, and
this conclusion has significant effects on parameter estimation of various
cosmological models \cite{WLZ14,WWGZ14,WWZ14,Wang15,WWL2016}. In addition, different
light-curve fitters of SNIa can also affect the results of cosmology-fits
\cite{Bengochea11,Bengochea14,HLLW15}. But in this work, for simplicity, we just
adopt the most mainstream recipe of processing SNLS3 data, and do not consider
the factors of time-varying $\beta$ and different light-curve fitters.}.

For comparison, we also use the JLA supernova samples (hereafter JLA)
\cite{Betoule2014}.

\item
The BAO data we used is extracted from three BAO measurements: the Baryon
Oscillation Spectroscopic Survey Data Release 9(BOSS DR9) \cite{Eisenstein2011},
the Sloan Digital Sky Survey Data Release 7(SDSS DR7) \cite{Abazajian2009} and
the 6dF Galaxy Survey (6dFGS) \cite{Beutler2011}.
There are two kinds of BAO data:
the first is extracted by using the spherically averaged one-dimensional galaxy clustering statistics (hereafter ``BAO1d''),
while the second is obtained by using the anisotropic two-dimensional GC statistics (hereafter ``BAO2d'') \cite{HLLW15bao}.

BAO1d includes the measurement of $r_s(z_d)/D_v(z=0.106)$ from \cite{Beutler2011},
$D_v(z=0.35)/r_s(z_d)$ from \cite{Abazajian2009}, and $D_v(z=0.57)/r_s(z_d)$ from \cite{Eisenstein2011}. Here $r_s(z_d)$ is the comoving sound horizon
at ``drag'' epoch and $D_v(z)$ is a volume averaged distance indicator (the corresponding
expressions are given in \cite{Wangyun2014}). For
convenience, we define:
\begin{eqnarray}
  \mbox{\bf q} &\equiv &(r_s(z_d)/D_v(z=0.106), D_v(z=0.35)/r_s(z_d), \nonumber\\
 && D_v(z=0.35)/r_s(z_d)).
\end{eqnarray}
Then we can represent the $\chi^2$ function of BAO1d dataset as
\begin{equation}
  \chi^2_{BAO1d} = \sum_{i} \frac{(q_i-q_i^{data})^2}{\sigma(q_i)^2},
\end{equation}
where $q_i^{data}$ and $\sigma(q_i)$ are the observed value and the
$1\sigma$ error of $q_i$.

BAO2d includes the measurement of $r_s(z_d)/D_v(z=0.106)$ from \cite{Beutler2011},
the measurements of $H(z=0.35)r_s(z_d)/c$ and $D_A(z=0.35)/r_s(z_d)$ from \citep{Hemantha2014},
as well as the measurements of $H(z=0.57)r_s(z_d)/c$ and $D_A(z=0.57)/r_s(z_d)$ from \cite{Wangyun2014}.
For convenience, we define:
\begin{eqnarray}
  \mbox{\bf p} & \equiv &(r_s(z_d)/D_v(z=0.106), H(z=0.35)r_s(z_d)/c, \nonumber\\
 && D_A(z=0.35)/r_s(z_d), H(z=0.57)r_s(z_d)/c, \nonumber\\
  &&D_A(z=0.57)/r_s(z_d)).
\end{eqnarray}
Then we can represent the $\chi^2$ function of BAO2d dataset as
\begin{equation}
  \chi^2_{BAO2d} = \sum_{i,j} (p_i-p_i^{data}) (Cov_{BAO2d}^{-1})_{ij} (p_j-p_j^{data}) ,
\end{equation}
where $p_i^{data}$ is the observed value of $p_i$ and $Cov_{BAO2d}$ is the
covariance matrix. For the details of the covariance matrix $Cov_{BAO2d}$, we
refer the reader to the Refs. \cite{Hemantha2014,Wangyun2014}.

\item
The cosmic microwave background data we used is the distance priors data
extracted from Planck 2015 data release (hereafter Planck 2015)
\cite{Planck201514} \footnote{In addition to \cite{Planck201514}, there are some
other distance priors data extracted from the Planck 2015 data release, i.e. see
Refs \cite{WangDai15,Huang2015,WangWang2013b}.}. This dataset use three quantities $l_a$, $R$ and
$\Omega_{b0} h^2$ ($h \equiv H_0/100$) to provide an efficient summary of CMB data.
Here $l_a$, $R$ and $\Omega_{b0} h^2$ are the acoustic scale, the CMB shift
parameter and the present baryon component, respectively. All these quantities
are defined as follows:
\begin{eqnarray}
l_a &=& \pi r(z_{\ast})/r_s(z_{\ast}) \\
R  &=& \sqrt{\Omega_{m0} H_0^2}r(z_{\ast})/c \\
\Omega_{b0} &=& \rho_{b0} /(3M_p^2H_0^2)
\end{eqnarray}
where $z_{\ast}$ is the redshift to the photon-decoupling surface given
in Ref. \citep{HuSugiyama1996},
$r(z_{\ast})$ is the comoving distance to $z_{\ast}$, and $r_s(z_{\ast})$
is the comoving sound horizon at $z_{\ast}$ (the corresponding
expressions are given in \cite{WangWang2013b}). For convenience, let's define:
$\mbox{\bf par} \equiv (l_a, R, \Omega_{b0} h^2)$. Then the $\chi^2$ function
can be written as
\begin{equation}
  \chi^2_{CMB} = \sum_{i,j} (par_i-par_i^{data}) (Cov_{CMB}^{-1})_{ij} (par_j-par_j^{data}) ,
\end{equation}
where $par_i^{data}$ is the observed value of $par_i$ and $Cov_{CMB}$ is the
covariance matrix, which is given in the Refs. \cite{Planck201514}.

For comparison, we also use the distance priors data extracted from Planck 2013
data release (hereafter Planck 2013) \cite{WangWang2013b}.

\item
The linear perturbation theory tell us
\begin{equation}
  \ddot{\delta_m} + 2 H \dot{\delta_m} - 4\pi G\rho_m \delta_m = 0,
  \label{eq:delta_m}
\end{equation}
where $\delta_m \equiv \delta \rho_m / \rho_m$ is the matter density perturbation.
Assuming that $D(z)$ is a solution of Eq. \ref{eq:delta_m}, it is clear that $D(0)=1$ and $D(\infty)=0$.
Therefore, the growth rate of large scale structure is $f(z)=-dlnD/dln(1+z)$,
and the root-mean-square mass fluctuation in $8h^{-1}$ Mpc spheres is $\sigma_8(z)=\sigma_8^0 D(z)$,
where $\sigma_8^0$ is the current value of $\sigma_8(z)$.
So we can get \cite{Pavlov2014}
\begin{equation}
\chi_{g}^2(\sigma_8^0) = \sum_{i=1}^N \frac{A(z_i,\sigma_8^0)-A_{data}(z_i)]^2}
{\sigma_i^2},
\end{equation}
where $N$ is the number of data points, $A(z,\sigma_8^0) \equiv f(z) \sigma_8(z)$
is the growth parameter, $A_{data}(z_i)$ and $\sigma_i$ are the mean value and 1$\sigma$
error of $A$. All the GF data can be obtained from the first table of \citep{Pavlov2014}.
Moreover, the posterior probability density function $\mathcal{L}_{g}$ is given by \cite{Pavlov2014}
\begin{equation}
\label{Lgrowth_Marg}
\mathcal{L}_{g} = \frac{1}{\sigma_{\overline{\sigma_8^0}}\sqrt{2\pi}}
\int_{0}^{\infty} {\rm exp}\left\lbrace-\frac{\chi^2_{g}(\sigma_8^0)}{2}-\frac{\left[ \sigma_8^0-\overline{\sigma_8^0} \right]^2}{2\sigma^{2}_{\overline{\sigma_8^0}}} \right\rbrace \mathrm{d}\sigma_8^0.
\end{equation}
where $\overline{\sigma_8^0} = 0.813(\Omega_{m0}/0.25)^{-0.47}$ is the mean value of $\sigma_8^0$,
and $\sigma_{\overline{\sigma_8^0}}=\sqrt{\sigma^2_{\sigma_8^0}+b^2}(\Omega_{m0}/0.25)^{-0.47}$
is the $1\sigma$ uncertainty of $\sigma_8^0$.
The final $\chi^2$ function of GF data can be written as
\begin{equation}
\chi^2_{g} = -2 \ln{{\mathcal L}_{g}}.
\end{equation}
\end{itemize}

In the original paper \cite{Hu2015}, the authors choose
\begin{equation}
  {\bf P}=\{\Omega_{m0}h^2, \Omega_{b0}h^2, h, C, \Omega_{\Lambda0},
\Omega_{k0}, \alpha, \beta\}
\end{equation}
as a set of free parameters to perform an MCMC likelihood analysis
\footnote{$\Omega_{r0}$ can be calculated by $\Omega_{r0} = \Omega_{\gamma 0}(1+0.2271 N_{eff})$,
where $\Omega_{\gamma 0} = 2.469 \times 10^{-5} h^{-2}$ and $N_{eff}=3.046$.}.
However, this choice will lead to the result of $\Omega_{hde0}$
has a negative 2$\sigma$ lower bound, which is unphysical. So in this work we
choose
\begin{equation}
  {\bf P}=\{\Omega_{b0}, h, C, \Omega_{hde0}, \Omega_{de0},
  \Omega_{k0}, \alpha, \beta\}
\end{equation}
as a set of free parameters. Moreover, we require that $\Omega_{hde0}>0$ is always satisfied.
Note that $\Omega_{de0}=\Omega_{hde0}+\Omega_{\Lambda 0}$ is the total fractional DE density of today.

\section{Results}
\label{sec:results}

In this section, First of all, we make use of the SNLS3+BAO1d+Planck 2015+GF data to constrain the $\Lambda$HDE model,
and compare its cosmology-fit results with the results of the $\Lambda$CDM and the HDE model.
Then, we discuss the impact of curvature on the cosmic evolutions of $\Omega_{\Lambda}$ and $\Omega_{hde}$
from both the theoretical and the observational aspects.
Finally, we explore the effects of adopting different types of observational data on the cosmic evolutions and the cosmic fate.

\subsection{A comparison of the cosmology-fit results of different DE models}
\label{subsec:mc}

\begin{table}\small
  \caption{Cosmology-fit results, $\chi^2_{min}/dof$, AIC and BIC of the $\Lambda$CDM,
    the HDE and the $\Lambda$HDE model.
    The SNLS3+BAO1d+Planck 2015+GF data are used in the analysis.
    Both the best-fit values and the $1\sigma$ errors of various parameters are listed.}
  \label{tab:fitting_result_m}
  \begin{tabular}{p{1.2cm}p{1.9cm}p{1.9cm}p{1.9cm}} 
          \hline	\hline
          Parameter & $\Lambda$CDM & HDE & $\Lambda$HDE \\
          \hline
          $\alpha$  & $1.41^{+0.30}_{-0.24}$ & $1.45^{+0.29}_{-0.31}$  & $1.41^{+0.35}_{-0.27}$ \\
          $\beta$ & $3.24^{+0.29}_{-0.26}$ & $3.25^{+0.32}_{-0.29}$ & $3.27^{+0.29}_{-0.31}$ \\
          $\Omega_{k0}$ & $-0.001^{+0.007}_{-0.008}$ & $0.003^{+0.011}_{-0.008}$ & $-0.006^{+0.013}_{-0.012}$ \\
          $\Omega_{b0}$ & $0.048^{+0.003}_{-0.004}$ & $0.046^{+0.005}_{-0.005}$ & $0.047^{+0.005}_{-0.005}$ \\
          $h$ & $0.689^{+0.025}_{-0.025}$ & $0.701^{+0.036}_{-0.036}$ & $0.693^{+0.040}_{-0.034}$ \\
          C &                   & $0.661^{+0.233}_{-0.125}$ & $0.334^{+2.666}_{-0.333}$ \\
          $\Omega_{hde0}$ &          & $0.710^{+0.028}_{-0.036}$ & $0.220^{+0.386}_{-0.220}$ \\
          $\Omega_{de0}$ & $0.704^{+0.021}_{-0.023}$ & $0.710^{+0.028}_{-0.036}$ & $0.713^{+0.031}_{-0.033}$  \\
          \hline
          \hline
          $\chi^2_{min}/dof$ & $0.912$ & $0.914$ & $0.913$ \\
         $\Delta$ AIC  & $0$ & $1.955$ & $2.574$ \\
         $\Delta$ BIC  & $0$ & $6.127$ & $10.917$ \\
          \hline
	\end{tabular}
\end{table}

By using the SNLS3+BAO1d+Planck 2015+GF data, we present the cosmology-fit
results of the $\Lambda$HDE model in the table \ref{tab:fitting_result_m}.
For comparison, the cosmology-fit results of the $\Lambda$CDM model and the HDE model
are also listed. We find that the results of $\Omega_{k0}$ of the three DE models
are consistent with the result of a flat universe at $1\sigma$ confidence
level (CL), which are also consistent with the result of \cite{Planck201513}.
In addition, the best-fit value of $C$ of the $\Lambda$HDE model is smaller
than that of the HDE model. Note that $C>1$ corresponds to a quintessence type
HDE, while $C<1$ corresponds to a phantom type HDE \cite{Li2004}. Therefore,
for the $\Lambda$HDE model, the HDE component is closer to a phantom DE
than that of the original HDE model. Moreover, from this table we see that, the
results of $\Omega_{m0}$ and $\Omega_{b0}$ of these three DE models are very
close. This means that, although these three DE models have very
different DE components, they indeed have a similar total fractional dark energy
density, i.e. $\Omega_{de0}\sim 0.7$. Moreover, to assess these three DE
models, we list the $\chi^2_{min}/dof$, the Akaike information criteria (AIC)
\cite{Akaike1974} and the Bayesian information criteria (BIC) \cite{Schwarz1978}
of three DE models in the table \ref{tab:fitting_result_m}. The AIC and the BIC
are defined as:
\begin{equation}
  \mathrm{AIC} = \chi^2_{min} + 2k, \mathrm{BIC} = \chi^2_{min} + k \ln{N}
\end{equation}
where $k$ is the number of free parameters, and $N$ is the number of data points.
From the table \ref{tab:fitting_result_m}, we find that, all the three criterias
(i.e. $\chi^2_{min}/dof$, AIC and BIC) indicate that the $\Lambda$CDM model gives
the best cosmology-fits among the three DE models. In other words, from the sight of model
fitting, the $\Lambda$CDM model is the best DE model of three DE models. However,
there are some other problems in the $\Lambda$CDM model, such as the
cosmological coincidence problem. Moreover, this problem can be solved in the framework
of the $\Lambda$HDE model (see Appendix \ref{ccp}). Therefore, it is necessary
to consider the case of beyond the standard cosmological model.

Therefore, we can conclude that making use of the cosmology-fit results given by the current observations
cannot distinguish the $\Lambda$HDE model from the $\Lambda$CDM and the HDE model.
In other words, even if we can obtain the exact value of $\Omega_{de0}$,
we still cannot determine the specific composition of DE.
This indicates that, the possibility that DE may contain multiple components cannot be rule out by the cosmological observations.

\subsection{The impact of curvature on the cosmic evolutions and the cosmic fates of the $\Lambda$HDE model}

In this subsection, we investigate the impact of curvature on
the cosmic evolutions and the corresponding cosmic fates of the $\Lambda$HDE model from both the theoretical and the
observational aspects. It should be mentioned that this topic has not been studied in the past.

Let us start from the theoretical side. From the Friedmann equation \ref{f.e.} we can derive
\begin{equation}
(1-\Omega_{hde})H^2=\Omega_{m0}H_0^2a^{-3}+\Omega_{r0}H_0^2a^{-4}+\Omega_{k0}H^2_0a^{-2}+\Omega_{\Lambda 0}H_0^2,
\end{equation}
where $a$ is the scale factor.
Defining a new function
\begin{equation}
f(a) \equiv \Omega_{m0}a^{-1}+\Omega_{r0}a^{-2}+\Omega_{k0}+\Omega_{\Lambda 0}a^2,
\end{equation}
we can obtain (see Appendix \ref{app})
\begin{equation}~\label{omega_hde1}
a\frac{d\Omega_{hde}}{da}=[\frac{2}{C}\sqrt{\Omega_{hde}+C^2 \Omega_k}-a\frac{d}{da}\ln{\left|f(a)\right|}]\Omega_{hde}(1-\Omega_{hde}).
\end{equation}
It has been proved that, for the $\Lambda$HDE model, the scale factor $a$ can eventually approach
infinity in a flat universe \cite{Hu2015}. In this work, we prove that this conclusion can be
extended to the case of a non-flat universe (see Appendix \ref{app}). Then, when $a$ is large enough,
the following approximation condition
\begin{equation}
  a\frac{d}{da}\ln{\left|f(a)\right|}=\frac{2\Omega_{\Lambda 0}a^2-\Omega_{m0}a^{-1}-2\Omega_{r0}a^{-2}}
  {\Omega_{\Lambda 0}a^2+\Omega_{m0}a^{-1}+\Omega_{r0}a^{-2}+\Omega_{k0}} \approx 2
  \label{condition}
\end{equation}
will be satisfied. In addition, for this case, the term of $\Omega_{k}$ can be also neglected.
Therefore, we can obtain the following equation
\begin{equation}
a\frac{d}{da}\ln{\left|\frac{\Omega_{hde}}{1-\Omega_{hde}}\right|} = \frac{2}{C}\sqrt{\Omega_{hde}} - 2.
\label{maink1}
\end{equation}
Notice that Eq. \ref{maink1} has the same form with the case of a flat universe.
This equation has an asymptotic solution
\begin{eqnarray}
  \ln{a}+x_{1} &=& \frac{C}{2C-2} \ln{\left| 1-\sqrt{\Omega_{hde}} \right|}-
  \frac{1}{2}\ln{\Omega_{hde}} \nonumber \\
  & & -\frac{1}{(C^2-1)} \ln{\left|\sqrt{\Omega_{hde}}-C \right|} \nonumber \\
  & & +\frac{C}{2C+2}\ln{(1+\sqrt{\Omega_{hde}})}, \label{mainsol2}
\end{eqnarray}
where $x_1$ is the constant of integration. This solution is also
the same as the result obtained in a flat universe \cite{Hu2015}.
Therefore, we can conclude that the asymptotic solution of $\Omega_{hde}$
obtained in a flat universe can be extended to the case of a non-flat universe.

\begin{table*}
  \caption{The finally evolution trend of $\Omega_{hde}$ and the corresponding cosmic fate for
  different initial conditions. The initial conditions are
  listed in the first column. The corresponding evolution trends of
  $\Omega_{hde}$ are given in the second column. The final fates of the universe
  are presented in the last column.
}
  \label{tab:fate}
  \begin{tabular}{ccc} 
          \hline
          \hline
          Initial condition & evolution trend &fate of universe \\
          \hline
          $\Omega_{hde}<1<C^2$ & $\Omega_{hde}$ will decrease to $0$ & eternal accelerated expansion\\
          $\Omega_{hde}<C^2<1$ & $\Omega_{hde}$ will decrease to $0$ & eternal accelerated expansion\\
          $C^2<\Omega_{hde}<1$ & $\Omega_{hde}$ will increase to $1$ & big rip \\
          $C^2<1<\Omega_{hde}$ & $\Omega_{hde}$ will decrease to $1$ & big rip \\
          $1<\Omega_{hde}<C^2$ & $\Omega_{hde}$ will increase to $C^2$ & eternal accelerated expansion\\
          $1<C^2<\Omega_{hde}$ & $\Omega_{hde}$ will decrease to $C^2$ & eternal accelerated expansion\\
         \hline
	\end{tabular}
\end{table*}

By analyzing the properties of this analytical solution,
in table \ref{tab:fate} we list the evolution trends of $\Omega_{hde}$ and the
corresponding cosmic fates given by the various initial conditions.
From this table we see that:
\begin{itemize}
\item
If $C^2>1$, then for the $\Lambda$HDE model, $\Omega_{hde}$ will eventually approach $0$ or $C^2$,
and the universe will undergo an eternal accelerated expansion.
\item
If $C^2<1$ and $\Omega_{hde}<C^2$, then for the $\Lambda$HDE model, $\Omega_{hde}$ will eventually approach $0$,
and the universe will also undergo an eternal accelerated expansion.
\item
If $C^2<1$ and $\Omega_{hde}>C^2$, then for the $\Lambda$HDE model, $\Omega_{hde}$ will eventually approach $1$,
and the universe will finally encounter a big rip.
\end{itemize}
These results are also consistent with that obtained in a flat universe \cite{Hu2015}.

\begin{figure*}
    \centering
      \resizebox{0.8\columnwidth}{!}{\includegraphics{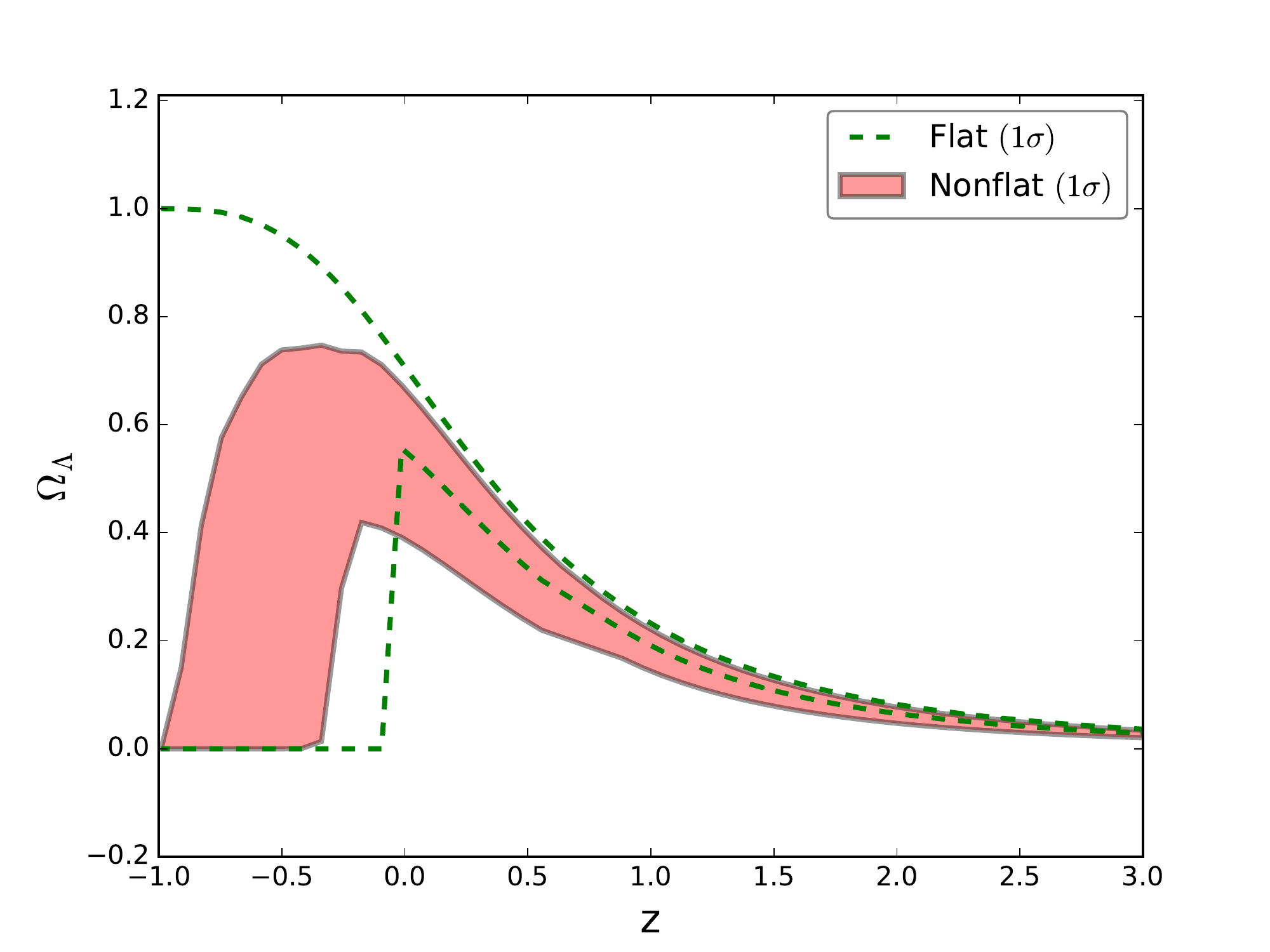}}
    \hspace{0.1\columnwidth}
      \resizebox{0.8\columnwidth}{!}{\includegraphics{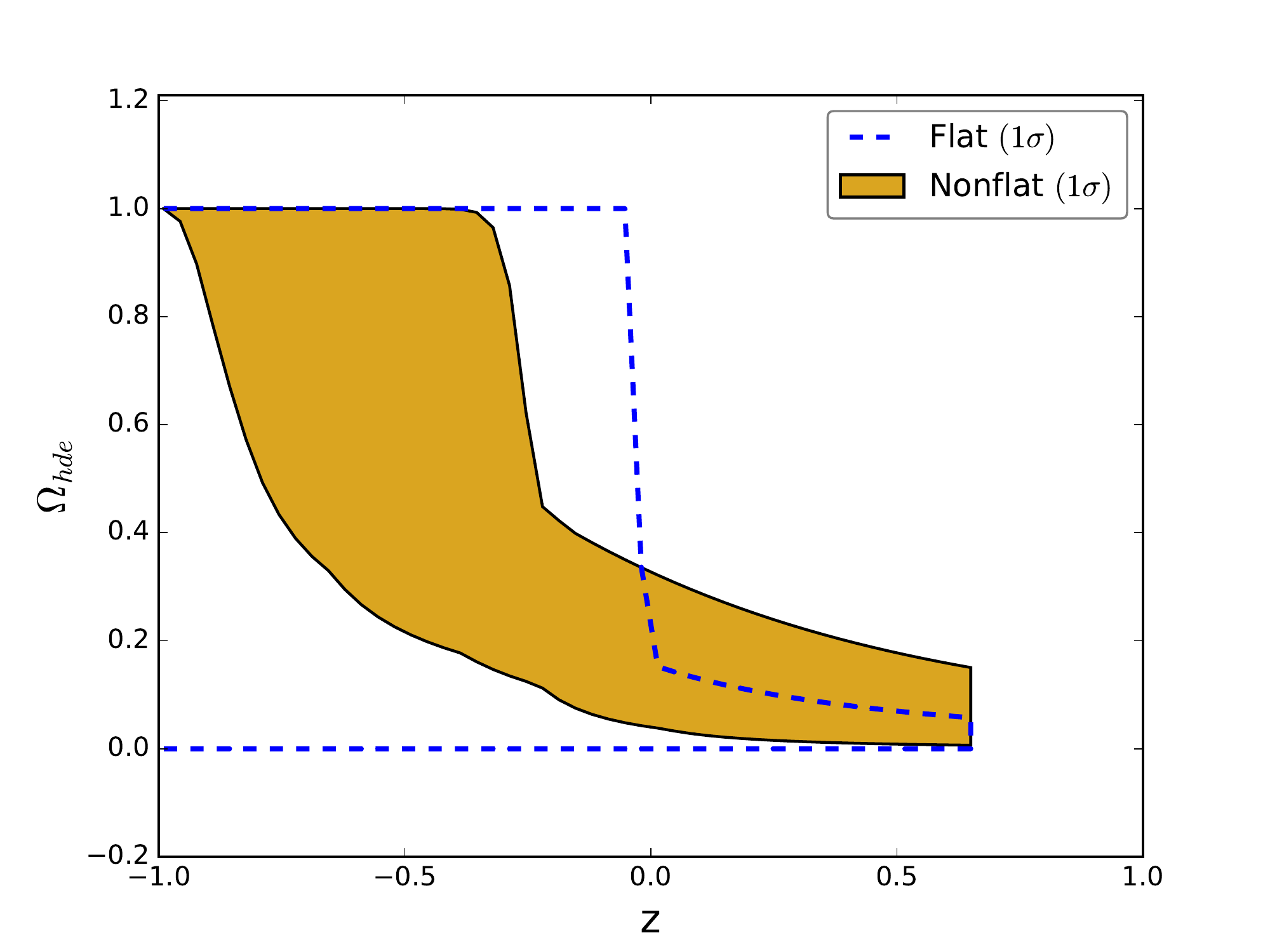}}
      \caption{
        The cosmic evolutions of $\Omega_{\Lambda}$ (left panel) and $\Omega_{hde}$ (right panel).
        In the left panel, the region inside green lines and the pink filled region
        denote the $1\sigma$ regions of $\Omega_{\Lambda}$ in a flat universe and a non-flat universe, respectively;
        In the right panel, the region inside blue lines and the gold filled region
        denote the $1\sigma$ regions of $\Omega_{hde}$ in a flat universe and a non-flat universe, respectively.
      }
      \label{fig:cf}
\end{figure*}

From the above discussions, we can see that only performing a qualitative theoretical analysis is
not enough to reflect the differences between a flat universe and a non-flat universe.
For a further investigation, it is necessary to perform a quantitative numerical study.
So in the Fig.~\ref{fig:cf}, based on the cosmology-fit results given by the SNLS3+BAO1d+Planck 2015+GF data,
we reconstruct the cosmic evolutions of $\Omega_{\Lambda}$ (left panel) and $\Omega_{hde}$ (right panel) at $1\sigma$ confidence region.
It must be stressed that,
this choice of redshift region is quite different from the case of \cite{Hu2015}, in which the region [-1,0] is not considered.
This is because one of the main aims of the present work is to study the cosmic fate of the $\Lambda$HDE model,
which has not been considered in \cite{Hu2015}.
From this figure, we can see that there is significant differences between a flat universe and a non-flat universe.
For the flat case, when $z$ approaches $-1$, $\Omega_{\Lambda}$ may approach $0$ or $1$ (see the left panel);
correspondingly, $\Omega_{hde}$ may approach $1$ or $0$ (see the right panel).
On the other hand, for the non-flat case, $\Omega_{\Lambda}$ will approach $0$ (see the left panel);
correspondingly, $\Omega_{hde}$ will approach $1$ (see the right panel).
This means that, for this case the energy density of HDE will increase with time $t$, which is a typical feature of phantom DE.
Therefore, compared with the case of a flat universe, considering curvature will make HDE closer to a phantom DE.

\subsection{The effects of adopting different types of observational data on the cosmic evolutions and the
cosmic fates}
The discussions above only adopt the SNLS3+BAO1d+Planck 2015+GF data.
In this subsection, we explore the impact of adopting different types of observational data
on the cosmic evolutions and the cosmic fates of $\Lambda$HDE.
It should be pointed out that this topic has not been studied in the previous literatures.

\subsubsection{Impacts of adopting different SN data}

\begin{figure*}
    \centering
      \resizebox{1.0\columnwidth}{!}{\includegraphics{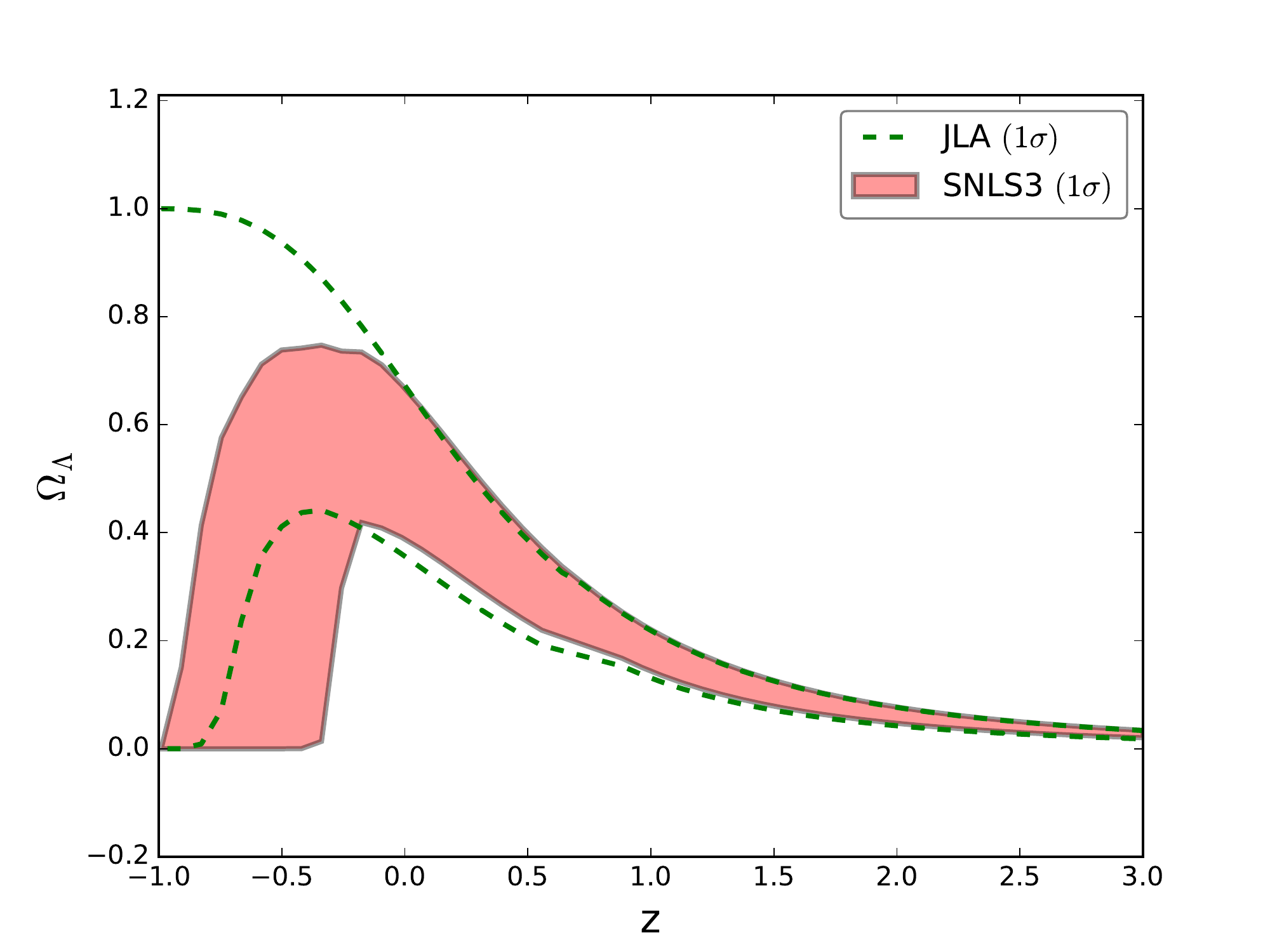}}
      \resizebox{1.0\columnwidth}{!}{\includegraphics{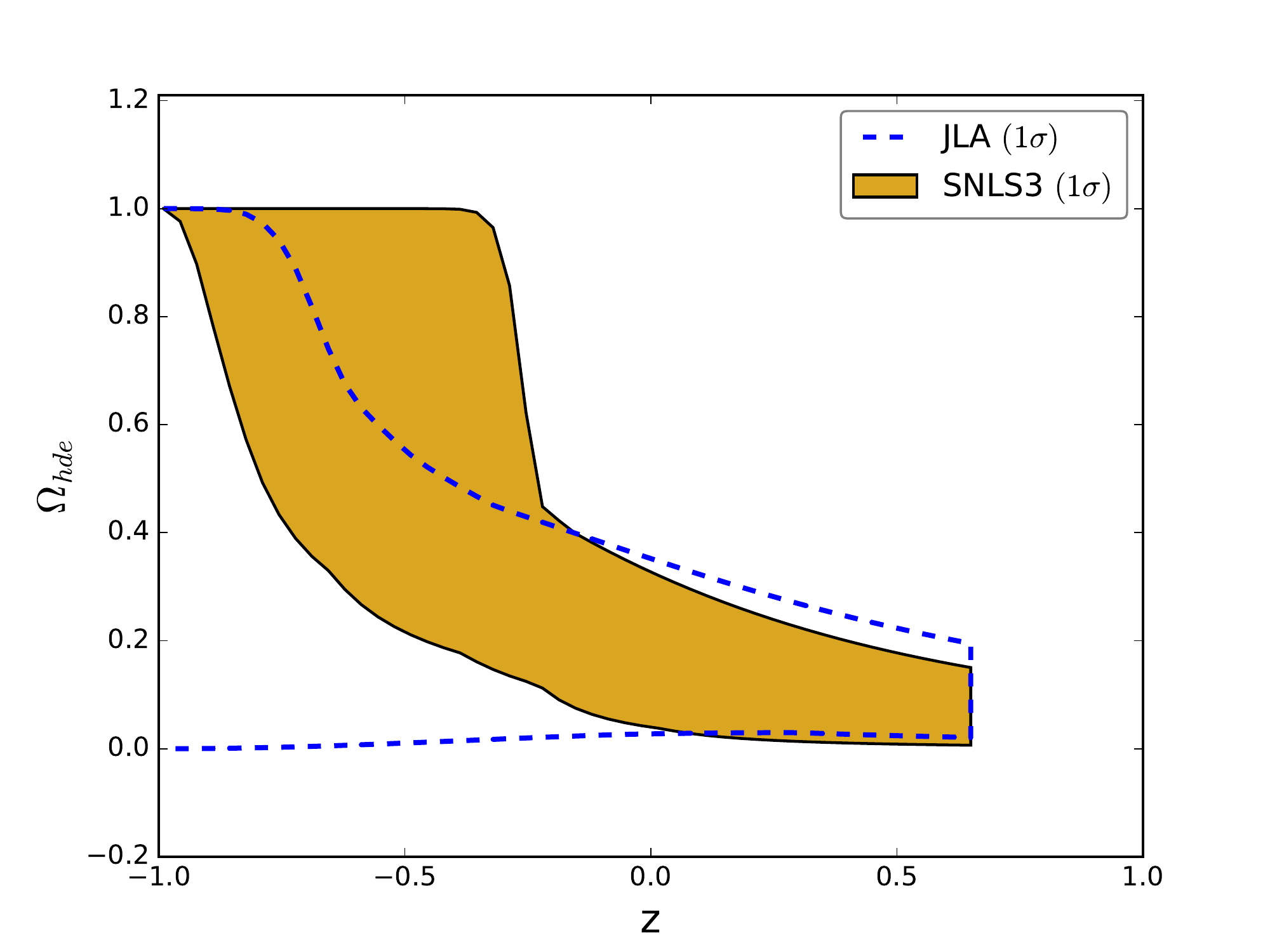}}
    \caption{
        The cosmic evolutions of $\Omega_{\Lambda}$ (left panel) and $\Omega_{hde}$ (right panel).
        In the left panel, the region inside green lines and the pink filled region
        denote the $1\sigma$ regions of $\Omega_{\Lambda}$ given by the JLA data and the SNLS3 data, respectively;
        In the right panel, the region inside blue lines and the gold filled region
        denote the $1\sigma$ regions of $\Omega_{hde}$ given by the JLA data and the SNLS3 data, respectively.
         ``JLA'' and``SNLS3''
         denote the JLA+BAO1d+Planck 2015+GF data and the SNLS3+BAO1d+Planck 2015+GF data, respectively.
      }
               \label{fig:SN}
\end{figure*}

Firstly, we study the impact of adopting different SN datasets.
For convenience, here we use ``SNLS3'' and ``JLA'' to represent
the SNLS3+BAO1d+Planck 2015+GF and the JLA+BAO1d+Planck 2015+GF datasets, respectively.
Making use of these two SN datasets, in Fig. \ref{fig:SN} we reconstruct
the evolutions of $\Omega_{\Lambda}$ (left panel) and $\Omega_{hde}$ (right panel) at $1\sigma$ confidence region.
From this figure we see that,
the $1\sigma$ region of $\Omega_{\Lambda}$ given by the ``JLA'' dataset has two possibilities:
$\Omega_{\Lambda}$ may eventually approach $1$ or $0$ (see the left panel);
correspondingly, $\Omega_{hde}$ may eventually approach $0$ or $1$ at $1\sigma$ CL (see the right panel).
In contrast, the $1\sigma$ region of $\Omega_{\Lambda}$ given by the ``SNLS3'' dataset has only one possibility:
$\Omega_{\Lambda}$ will eventually approach $0$ (see the left panel);
correspondingly, $\Omega_{hde}$ will eventually approach $1$ at $1\sigma$ CL (see the right panel).
This means that the adopting ``SNLS3'' data will yield a HDE dominated universe at $1\sigma$ CL.
In other words, compared with JLA dataset, SNLS3 dataset more favors a phantom type HDE.

\subsubsection{Impacts of adopting different BAO data}

\begin{figure*}
    \centering
      \resizebox{1.0\columnwidth}{!}{\includegraphics{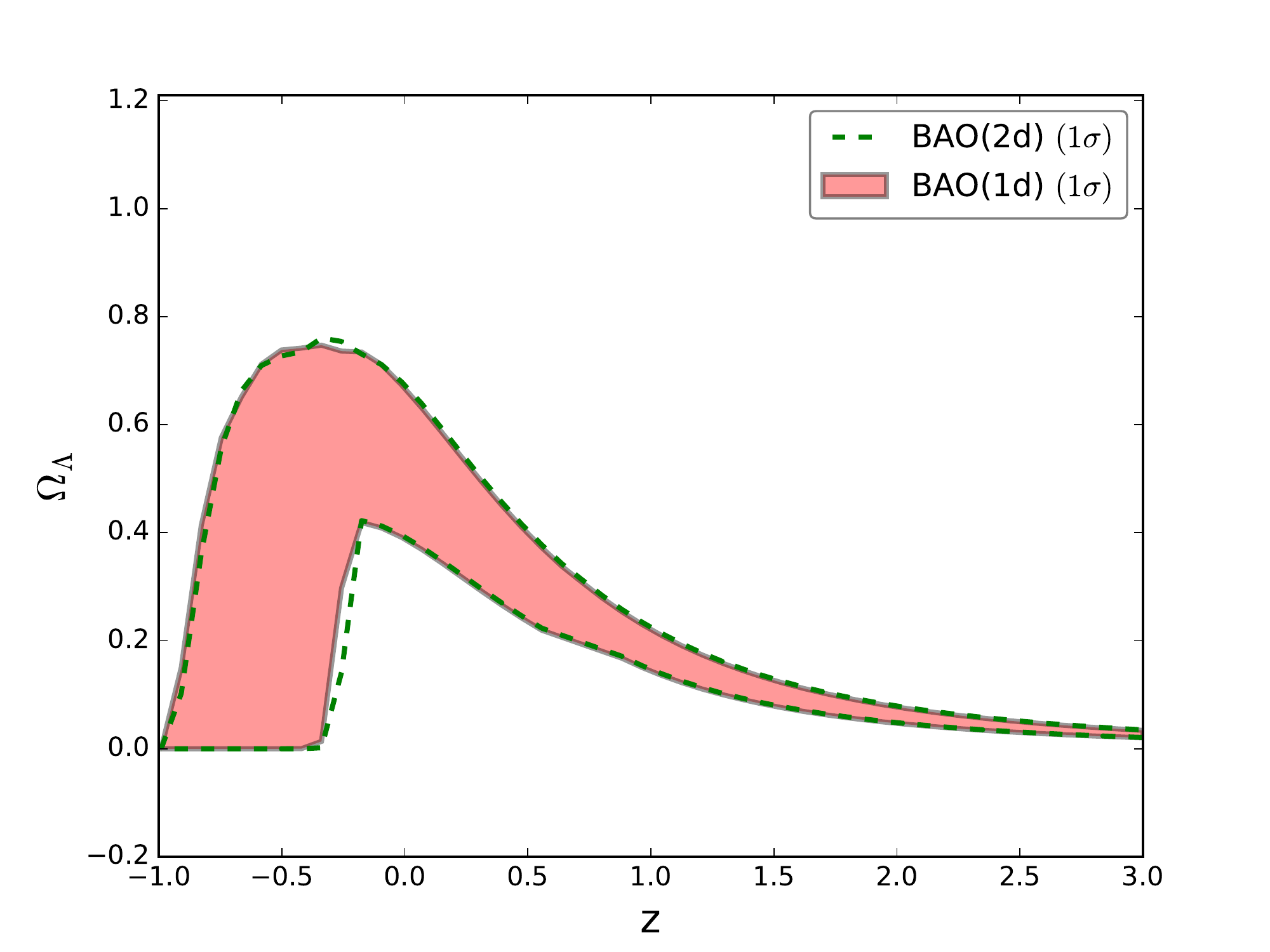}}
      \resizebox{1.0\columnwidth}{!}{\includegraphics{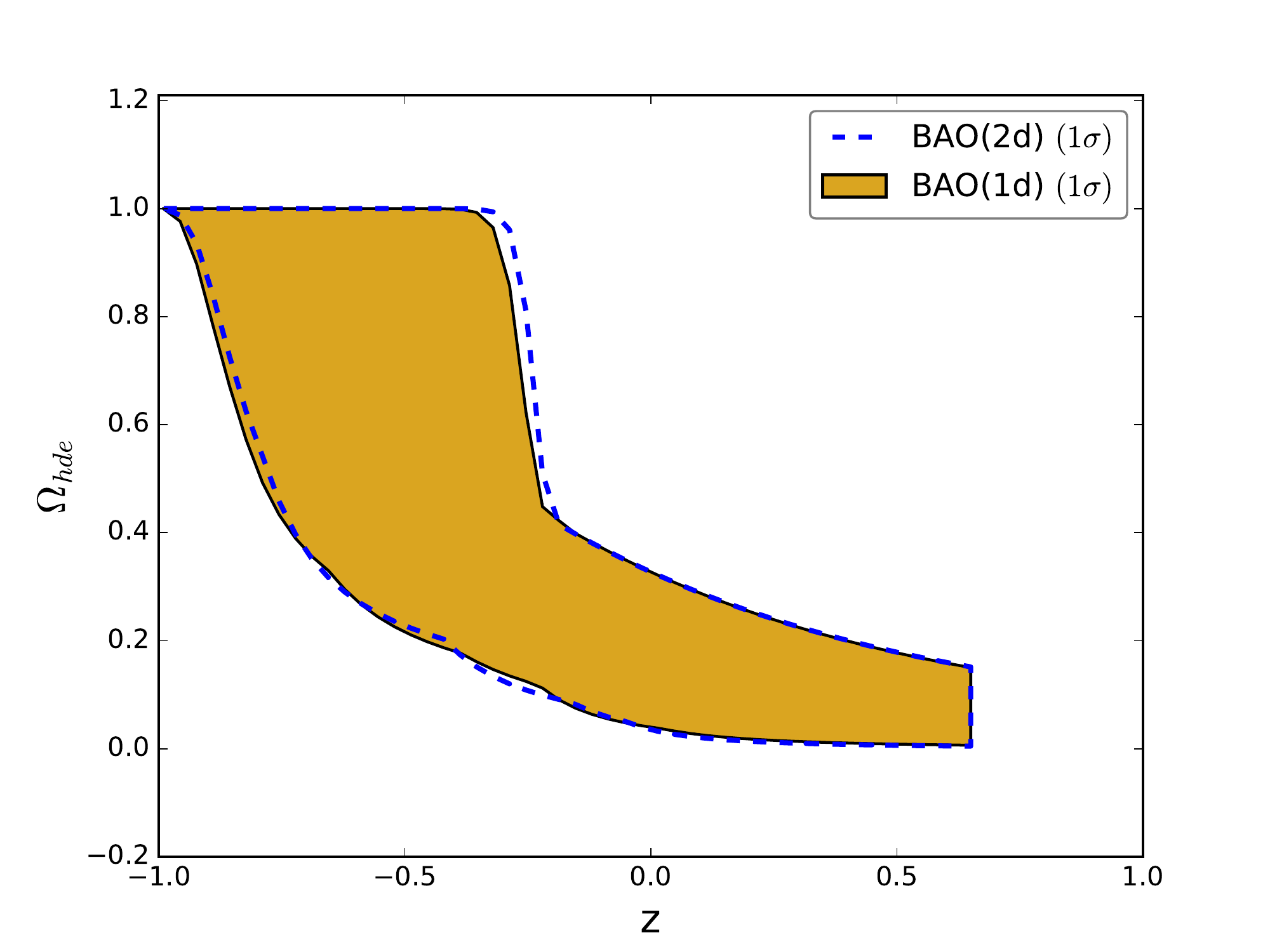}}
    \caption{
        The cosmic evolutions of $\Omega_{\Lambda}$ (left panel) and $\Omega_{hde}$ (right panel).
        In the left panel, the region inside green lines and the pink filled region
        denote the $1\sigma$ regions of $\Omega_{\Lambda}$ given by the BAO2d data and the BAO1d data, respectively;
        In the right panel, the region inside blue lines and the gold filled region
        denote the $1\sigma$ regions of $\Omega_{hde}$ given by the BAO2d data and the BAO1d data, respectively.
        ``BAO1d'' and ``BAO2d'' denote
        the SNLS3+BAO1d+Planck 2015+GF data and the SNLS3+BAO2d+Planck 2015+GF data, respectively.
      }
               \label{fig:BAO}
\end{figure*}

Next, we explore the effects of adopting different BAO data.
For convenience, here we
use ``BAO1d'' and ``BAO2d'' to represent the SNLS3+BAO1d+Planck 2015+GF
and the SNLS3+BAO2d+Planck 2015+GF datasets, respectively.
In Fig.~\ref{fig:BAO}, by using the ``BAO1d'' and the ``BAO2d'' datasets,
we reconstruct the evolutions of $\Omega_{\Lambda}$ (left panel) and $\Omega_{hde}$ (right panel) at $1\sigma$ confidence region.
It is clear that ``BAO1d'' dataset, which is indeed the same as the ``SNLS3'' dataset,
favors a HDE dominated universe at $1\sigma$ CL.
In addition, the evolutionary trajectory of $\Omega_{\Lambda}$ and $\Omega_{hde}$ given by the ``BAO2d'' dataset
almost overlap with the results of ``BAO1d'' dataset.
this means that using different BAO data have little impact on the cosmic evolutions of the $\Lambda$HDE model.

\subsubsection{Impacts of adopting different CMB data}

\begin{figure*}
    \centering
      \resizebox{1.0\columnwidth}{!}{\includegraphics{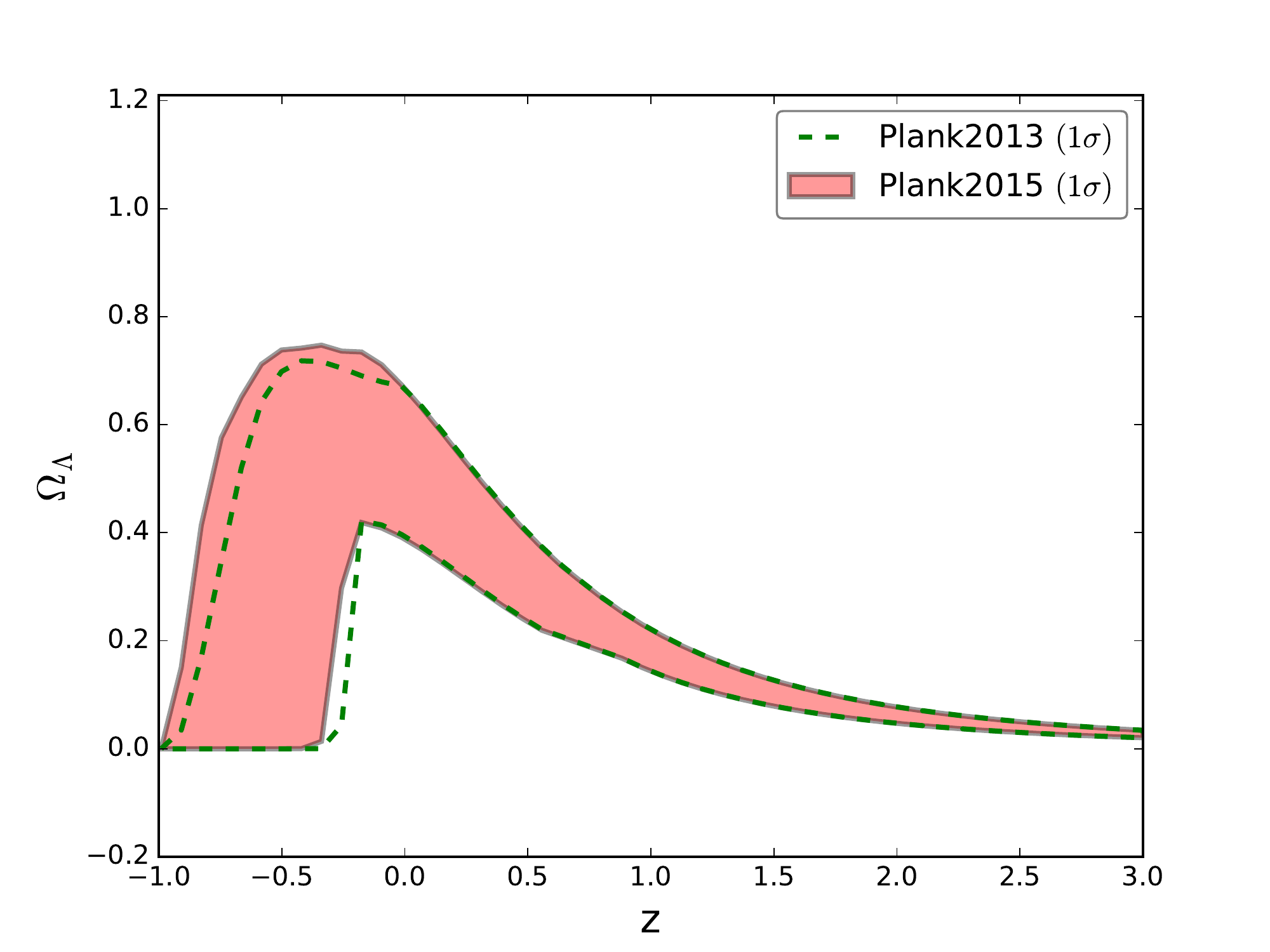}}
      \resizebox{1.0\columnwidth}{!}{\includegraphics{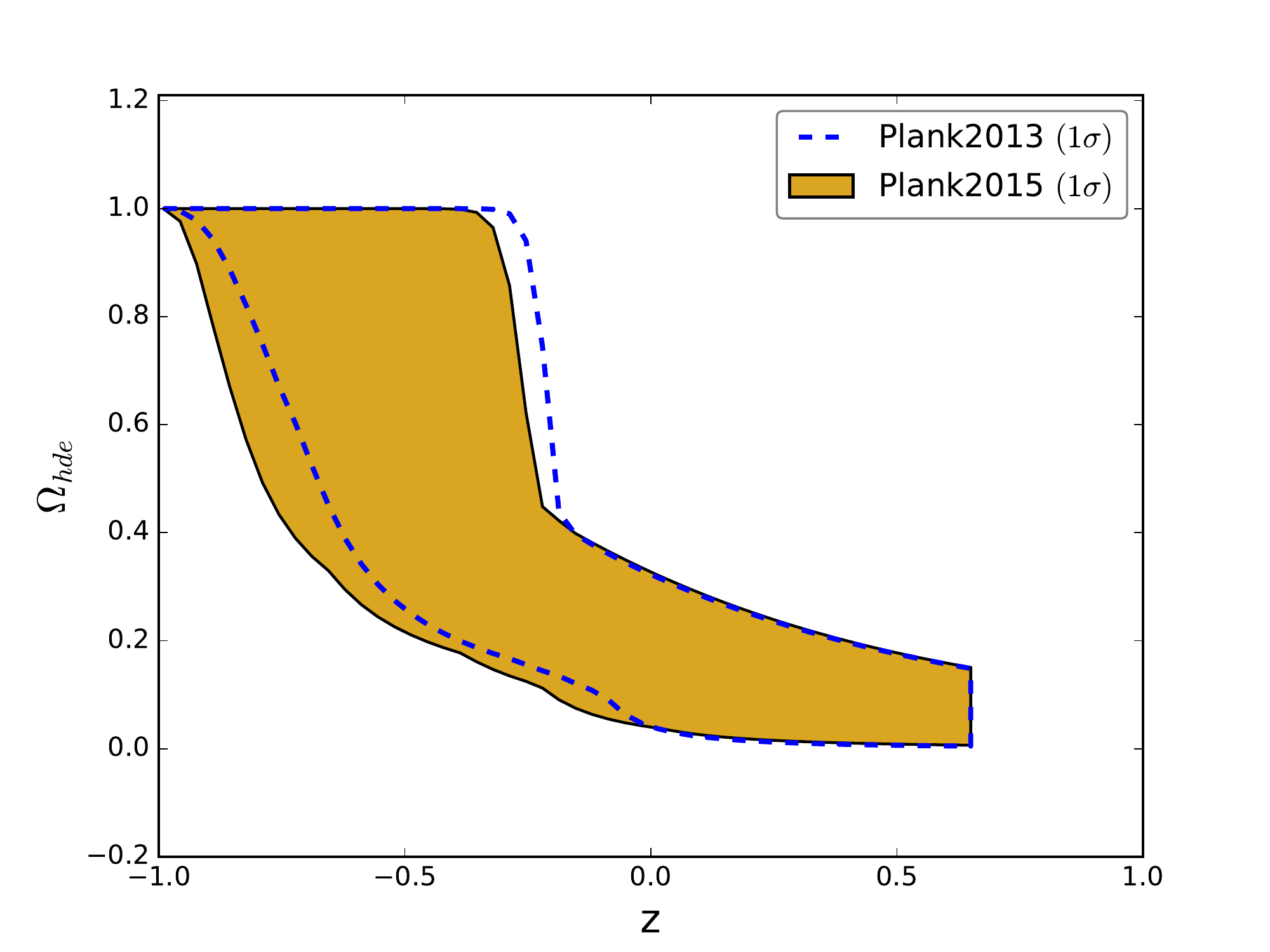}}
    \caption{
        The cosmic evolutions of $\Omega_{\Lambda}$ (left panel) and $\Omega_{hde}$ (right panel).
        In the left panel, the region inside green lines and the pink filled region
        denote the $1\sigma$ regions of $\Omega_{\Lambda}$ given by the Planck 2013 data and the Planck 2015 data, respectively;
        In the right panel, the region inside blue lines and the gold filled region
        denote the $1\sigma$ regions of $\Omega_{hde}$ given by the Planck 2013 data and the Planck 2015 data, respectively.
        ``Planck 2013'' and ``Planck 2015'' denote
        the SNLS3+BAO1d+Planck 2013+GF data and the SNLS3+BAO1d+Planck 2015+GF data, respectively.
      }
               \label{fig:CMB}
\end{figure*}

Then, we study the effects of adopting different CMB data.
Here we use ``Planck 2015'' and ``Planck 2013'' to represent
the SNLS3+BAO1d+Planck 2015+GF and the SNLS3+BAO1d+Planck 2013+GF datasets, respectively.
In Fig.~\ref{fig:CMB}, by adopting these two datasets, we reconstruct the evolutions of
$\Omega_{\Lambda}$ (left panel) and $\Omega_{hde}$ (right panel) at $1\sigma$ confidence region.
Again, the ``Planck 2015'' dataset, which is indeed the same as the ``SNLS3'' dataset,
favors a HDE dominated universe at $1\sigma$ CL.
In addition, the evolutionary trajectories of $\Omega_{\Lambda}$ and $\Omega_{hde}$ given by the ``Planck 2013'' dataset
almost overlap with the results of ``Planck 2015'' dataset.
this result is just the same as the case of using different BAO data.

\subsubsection{Impacts of adding growth factor data or not}

\begin{figure*}
    \centering
      \resizebox{1.0\columnwidth}{!}{\includegraphics{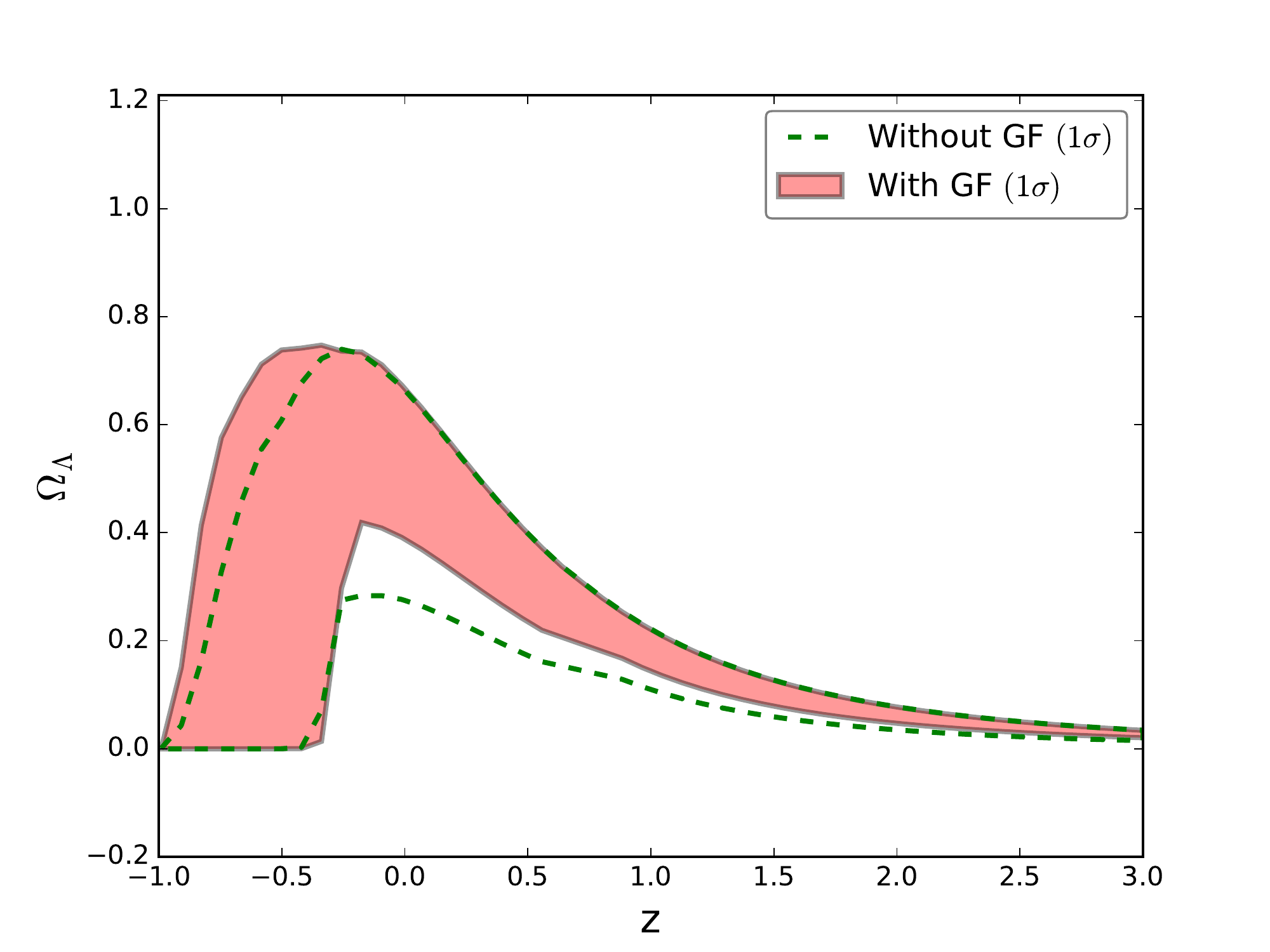}}
      \resizebox{1.0\columnwidth}{!}{\includegraphics{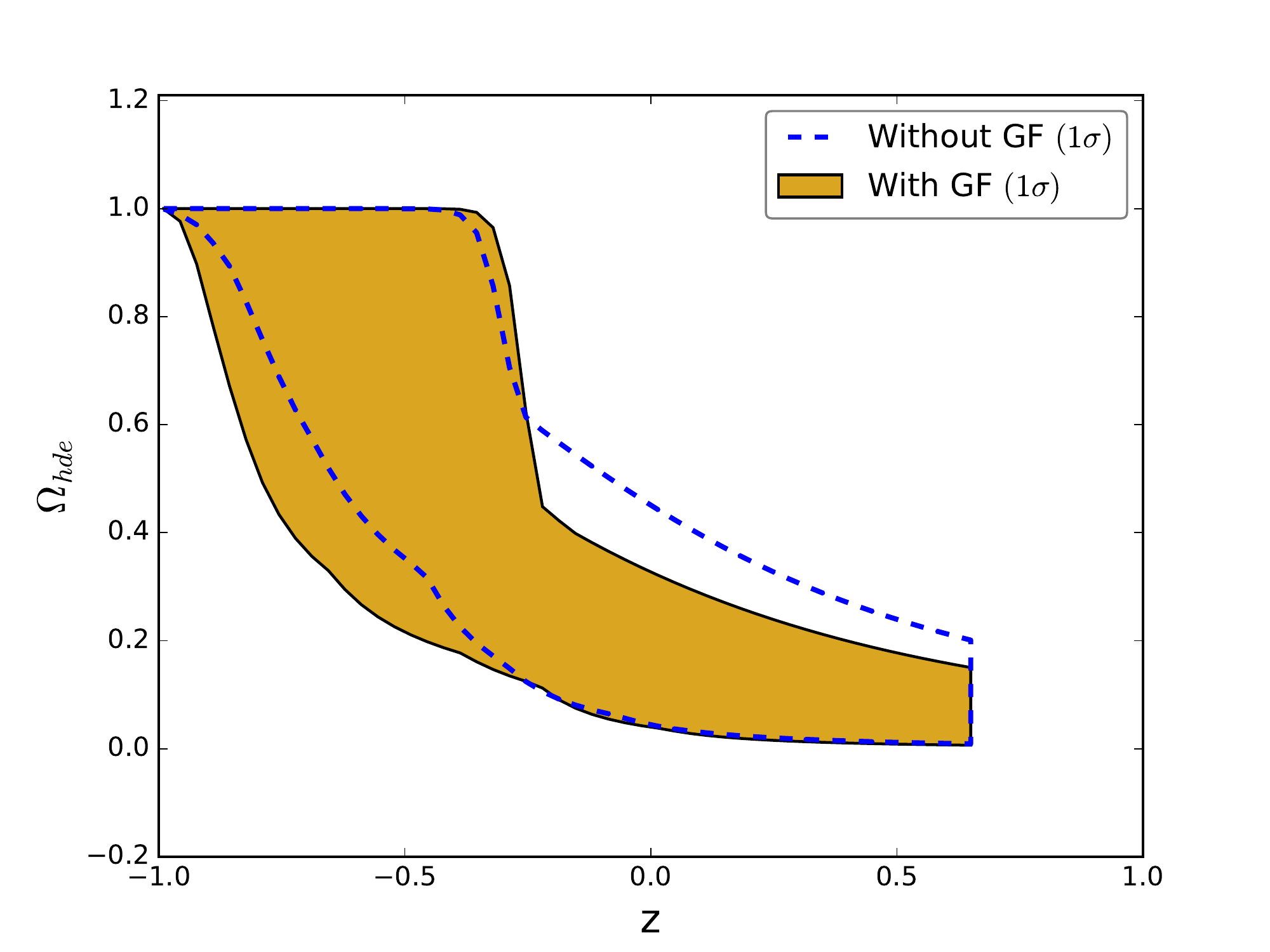}}
    \caption{
        The cosmic evolutions of $\Omega_{\Lambda}$ (left panel) and $\Omega_{hde}$ (right panel).
        In the left panel, the region inside green lines and the pink filled region
        denote the $1\sigma$ regions of $\Omega_{\Lambda}$ for the cases without and with GF data, respectively;
        In the right panel, the region inside blue lines and the gold filled region
        denote the $1\sigma$ regions of $\Omega_{hde}$ for the cases without and with GF data, respectively.
        ``Without GF'' and ``With GF''
        denote the SNLS3+BAO1d+Planck 2015 data and the SNLS3+BAO1d+Planck 2015+GF data, respectively.
      }
               \label{fig:GF}
\end{figure*}

Finally, we discuss the impact of adding GF data or not.
Here we use ``With GF'' and ``Without GF'' to represent the
SNLS3+BAO1d+Planck 2015+GF and the SNLS3+BAO1d+Planck 2015 datasets, respectively.
In Fig.~\ref{fig:GF}, by making use of these two datasets,
we reconstruct the evolutions of $\Omega_{\Lambda}$ (left panel) and $\Omega_{hde}$ (right panel) at $1\sigma$ confidence region.
From this figure we see that,
although there are small differences between the evolutionary trajectories given by the ``Without GF'' and the ``With GF'' datasets,
the overall evolutionary trends of the ``Without GF'' and the ``With GF'' cases are exactly the same.
Therefore, we can conclude that adding the GF data or not
will not have any significant impact on the cosmic evolutions of the $\Lambda$HDE model.

\section{Summary}
\label{sec:conclusion}

In the previous theoretical studies of DE, DE is always viewed as a single component.
So far as we know, the $\Lambda$HDE model, which is proposed in \cite{Hu2015},
is the first theoretical attempt to explore the possibility that DE contains multiple components.
In \cite{Hu2015}, we have performed some simply analyses on this model.
However, there are still some shortcomings for Ref. \cite{Hu2015}:
(1) the cosmology-fit results of the $\Lambda$HDE model have not been compared with the
results of other DE models;
(2) the impact of curvature on the cosmic evolutions and the cosmic fates of the $\Lambda$HDE model
have not been considered;
(3) the effects of adopting different types of observational data have not been taken into account.

The main aim of the present work is to give a more comprehensive and systematic investigation
on the cosmological implications of the $\Lambda$HDE model.
In particular, by combining the qualitative theoretical analyses with the quantitative numerical studies,
we explore in details the issues mentioned above.
It should be mentioned that, different from Ref. \cite{Hu2015}, in this work we study a new version of $\Lambda$HDE model:
in the deceleration expansion stage, DE only contains the CC term; in the accelerated expansion stage, DE contains both the CC and the HDE.
In addition, the observational data used in this work includes two types of SN data (SNLS3 and JLA), two types
of BAO data (BAO1d and BAO2d), two types of CMB data (Planck 2015 and Planck 2013) and the GF data.

Our conclusions are as follows:
\begin{itemize}
\item
The $\Lambda$HDE model has a similar value of $\Omega_{de0}$ with the $\Lambda$CDM and the HDE model(see Table \ref{tab:fitting_result_m});
this implies that making use of the current observations cannot distinguish the $\Lambda$HDE model from the other two models.
In other words, even if we can obtain the exact value of $\Omega_{de0}$,
we still cannot determine the specific composition of DE.
This indicates that, the possibility that DE may contain multiple components cannot be rule out by the cosmological observations.
\item
The qualitative theoretical analysis shows that, for the $\Lambda$HDE model, the asymptotic solution of $\Omega_{hde}$
and the corresponding cosmic fate (see Table \ref{tab:fate}) in a flat universe can be extended to the case of a non-flat universe.
On the other side, the quantitative numerical study shows that,
considering curvature will make HDE closer to a phantom dark energy. (see Fig. \ref{fig:cf}).
For this case, the universe will finally encounter a big rip
\item
compared with JLA dataset, SNLS3 dataset more favors a phantom type HDE (see Fig. \ref{fig:SN}).
In contrast, using other types of observational data have no significant impact on
the cosmic evolutions of $\Omega_{\Lambda}$ and $\Omega_{hde}$ (see Fig. \ref{fig:BAO}, Fig. \ref{fig:CMB} and Fig. \ref{fig:GF}).
These observational data all favor a phantom type HDE,
showing that the universe will finally encounter a big rip.
\end{itemize}

There are still many other topics, such as perturbation \cite{Li2008},
interaction \cite{LLWWZ09}, neutrino \cite{LWLZ13}, cosmic age \cite{LLLW10,WLL10}, standard siren \cite{Wang08c,Wangalone10} and so on,
deserve to be studied in the framework of the $\Lambda$HDE model.
In addition, it is also interesting to make use of various diagnosis tools
to analyze the cosmic evolution of the $\Lambda$HDE model \cite{ZW16}. These will be done in future work.

\section*{Acknowledgments}

We thank the referee for very helpful suggestions.
SW is supported by the National Natural Science Foundation of China under Grant
No. 11405024 and the Fundamental Research Funds for the Central Universities under Grant No. 16lgpy50.
ML is supported by the National Natural Science Foundation of China (Grant No. 11275247, and Grant No. 11335012)
and a 985 grant at Sun Yat-Sen University.

\

\appendix

\section{The solution of cosmological coincidence problem in the framework of
the $\Lambda$HDE model}
\label{ccp}
The cosmological coincidence problem was first proposed in a conference \cite{Steinhardt97}.
This problem can be expressed a problem of why
the radiation energy density is far greater than the dark energy density at the onset of the radiation dominated epoch.
Now, let us discuss the possible solution to the cosmological coincidence problem in the framework of the $\Lambda$HDE model.
Since the future event horizon does exist in the inflation stage, the HDE may also exist in this stage;
we can assume that, in the standard slow-roll inflation epoch there are two energy components:
the HDE and the inflation energy.
If the latter is almost constant,
we can derive the evolution equation of $\Omega_{hde}$:
\begin{equation}
a\frac{d}{da}\ln{\left|\frac{\Omega_{hde}}{1-\Omega_{hde}}\right|} = \frac{2}{C}\sqrt{\Omega_{hde}} - 2.
\label{inflation}
\end{equation}
By solving the above equation, we find that, when the scale factor $a$ is large
enough,
\begin{equation}
  \Omega_{hde} \sim a^{-2} .
\label{hde_inflation}
\end{equation}
From the above equation, we can see that, in the end of inflation,
$\Omega_{hde}$ will approach a small number. Meanwhile, we assume that most of
the inflation energy is decayed into radiation and the other part of the
inflation energy is included in a CC. Therefore, if the number of e-folds is
reasonable, we can get a very tiny ratio between the dark energy density and the
radiation density at the onset of the radiation dominated epoch. This means
that, the inflation can naturally solve the cosmological coincidence problem
without any fine-tuned initial conditions.

\section{A proof that scale factor $a$ has no maximum in the $\Lambda$HDE model}\label{app}
In this appendix, we prove that, in the framework of the $\Lambda$HDE model,
the scale factor $a$ does not have a maximum.

Firstly, we will proof that, if $a$ is finite, $\rho_{hde}=3C^2M_p^2/L^2\neq0$ is always satisfied.
We notice that the cutoff scale length $L$ can be written:
\be
L = a r(t),
\ee
where $r(t)$ is defined as
\be
r(t) = \frac{1}{\sqrt{|\Omega_{k0}|}H_0} sinn(\sqrt{|\Omega_{k0}|}H_0 \int_t^{+\infty} \frac{dt'}{a}).
\ee
The evolution of $r(t)$ has two possibilities:
\begin{itemize}
\item
If $\Omega_{k0}<0$, then $r(t)$ must be a bounded function,
because $sinn$ takes a $sin$ function form.
\item
If $\Omega_{k0} \geq 0$, then $r(t)$ must be a decreasing function.
For the $\Lambda$HDE model, the current HDE density $\rho_{hde0}>0$ is always satisfied, then $r(t_0)$ must be finite.
So in the future, $r(t)<r(t_0)$ is always satisfied.
\end{itemize}
The discussions above imply that $r(t)$ must be finite, which is independent of the specific value of $\Omega_{k0}$.
Therefore, when $a$ is finite, $r$ is also finite,
and $\rho_{hde}=3C^2M_p^2/(ar)^2 \neq 0$.

Secondly, we will proof that $a$ does not have a maximum.
Assuming that $a$ has a maximum $a_m$, then $H|_{a = a_m} = \frac{\dot{a}}{a}|_{a = a_m} = 0$.
Notice that $\rho_{hde}|_{a = a_m} \neq 0$, so
\be
\label{eq:Key1}
\Omega_{hde}|_{a = a_m}=\frac{\rho_{hde}}{3M_{pl}^2 H^2} \quad \bigg|_{a = a_m} \rightarrow \infty.
\ee

In the following, we will prove that the Eq. \ref{eq:Key1} contradict with the Friedmann
equation, which has the form
\begin{equation}
\label{eq:FE2} 3M_{pl}^2 H^2=\rho_{m}+\rho_r+\rho_k+\rho_{\Lambda}+\rho_{hde}.
\end{equation}
Let us define that
\begin{eqnarray}
  f(a)& \equiv & \Omega_{m0}a^{-1}+\Omega_{r0}a^{-2}+\Omega_{k0}+\Omega_{\Lambda 0}a^2,\\
  g(a)& \equiv &a\frac{d}{da}\ln{\left|f(a)\right|}.
\end{eqnarray}
It is easily to get
\be
\label{eq:ga}
g(a)=\frac{2\Omega_{\Lambda0}a^{2}-\Omega_{m0}a^{-1}-2\Omega_{r0}a^{-2}}{f(a)}.
\ee
Moreover, from the Friedmann equation \ref{eq:FE2}, we can derive
\begin{equation}
\label{omega_hde}
a\frac{d\Omega_{hde}}{da}=[\frac{2}{C}\sqrt{\Omega_{hde}+C^2 \Omega_k}-g(a)]\Omega_{hde}(1-\Omega_{hde}).
\end{equation}
Due to that the evolution of $\Omega_{hde}$ mainly
depends on the terms in the right hand of Eq. \ref{omega_hde},
we will discuss the values of these terms in the following.

Firstly, let's consider the function $g(a)$.
From the Eq. \ref{eq:FE2}, we get $(\rho_m + \rho_r + \rho_k + \rho_\Lambda)|_{a = a_m} = -\rho_{hde}|_{a = a_m} \neq 0$.
Then,
\be
f(a_m) \propto (\rho_m + \rho_r + \rho_k + \rho_\Lambda)|_{a = a_m} \neq 0.
\ee
Based on the Eq. \ref{eq:ga}, we can see that $g(a_m)$ must be finite.

Then, let us discuss other terms in Eq. \ref{omega_hde}.
When $a$ approaches $a_m$,
\begin{itemize}
\item
if $\Omega_{k}>0$, then $\sqrt{\Omega_{hde}+C^2 \Omega_k}$ will approach
infinity apparently;
\item
if $\Omega_{k}<0$, notice that the EoS of HDE satisfies \cite{Hu2015} $w_{hde}=-{1\over3}-{2\over3}\sqrt{{\Omega_{hde}\over C^2}+\Omega_k}<-\frac{1}{3}$,
we can get
\begin{eqnarray}
  \sqrt{\Omega_{hde}+C^2 \Omega_k} &=& \sqrt{\Omega_{hde}(1+C^2 \frac{\rho_{k0}}{\rho_{hde0}} a^{1+3 w_{hde}})} \nonumber \\
                 & > & \sqrt{\Omega_{hde}(1+C^2 \frac{\rho_{k0}}{\rho_{hde0}})}.
\end{eqnarray}
Since $\sqrt{\Omega_{hde}(1+C^2 \frac{\rho_{k0}}{\rho_{hde0}})}$ will approach infinity when $a \rightarrow a_m$,
$\sqrt{\Omega_{hde}+C^2 \Omega_k}$ will still approach infinity.
\end{itemize}
Therefore, the relation
\be
\sqrt{\Omega_{hde}+C^2 \Omega_k} \bigg|_{a = a_m} \rightarrow \infty
\ee
is also independent of the specific value of $\Omega_{k0}$.

Finally, let's discuss the evolution of $\Omega_{hde}$.
When $a$ approaches $a_m$, $\sqrt{\Omega_{hde}+C^2 \Omega_k}>\frac{C}{2}g(a)$
and $\Omega_{hde}>1$ will always be satisfied. So according to the Eq.(\ref{omega_hde}),
we can get
\begin{equation}
\label{eq:Key2}
\frac{d}{da}\sqrt{\Omega_{hde}}<0.
\end{equation}
It is clear that $\Omega_{hde}$ is a decreasing function of $a$,
and this means that $\Omega_{hde}$ will never approach infinity in the future.
This result apparently contradicts with the Eq. \ref{eq:Key1}.
So the previous assumption about $a$ is wrong, and $a$ has no maximum.


\label{lastpage}

\end{document}